\documentclass[prl,aps,amssymb,showpacs,twocolumn]{revtex4}
\usepackage{amsmath}
\usepackage{amssymb}
\usepackage{amsthm}
\usepackage{amsfonts}
\usepackage{enumerate}
\usepackage{latexsym}
\usepackage[dvips]{graphicx}

\newcommand{\beq}{\begin{equation}}
\newcommand{\eneq}{\end{equation}}

\input{epsf}

\begin{document}

\tolerance 10000

\newcommand{\vk}{{\bf k}}

%\draft

\title{Dissipationless Spin Current in Anisotropic p-Doped Semiconductors}
\author { Bogdan A. Bernevig$^\dagger$, JiangPing Hu$^\star$, Eran Mukamel$^\dagger$, Shou-Cheng
Zhang$^\dagger$}

\affiliation{ $\dagger$ Department of Physics, Stanford
University,
         Stanford, California 94305 \\
         $\star$ Department of Astronomy and Physics, UCLA, Los Angeles, California 90095}

\begin{abstract}
Recently, dissipationless spin current has been predicted for the
p-doped semiconductors with spin-orbit coupling. Here we
investigate the effect of spherical symmetry breaking on the
dissipationless spin current, and obtain values of the intrinsic
spin Hall conducitivity for realistic semiconductor band
structures with cubic symmetry.

\end{abstract}


\maketitle

Spintronics is a new field of science and technology which aims to
manipulate the spin of the electron for building functional logic
and storage devices~\cite{WOLF2001A}. The creation, manipulation
and transport of spin currents is a central challenge in this
field. Recently, Murakami, Nagaosa and Zhang\cite{MURAKAMI2003}
found a basic law of spintronics, which relates the spin current
and the electric field by the response equation
\begin{equation}
j_j^i = \sigma_s \epsilon_{ijk} E_k \label{spin_response}
\end{equation}
where $j_j^i$ is the current of the $i$-th component of the spin
along the direction $j$ and $\epsilon_{ijk}$ is the totally
antisymmetric tensor in three dimensions. This effect arises
because of the spin-orbit coupling in the valence band of
conventional semiconductors such as GaAs and Ge. Sinova et
al\cite{sinova2003} also found a similar effect in the electron
doped conduction band. The transport equation
(\ref{spin_response}) is similar to Ohm's law in electronics.
However, unlike the Ohm's law, this new law describes a purely
dissipationless spin current, in the sense that Eq.
(\ref{spin_response}) is invariant under the time reversal and the
intrinsic part of $\sigma_s$ does not depend on impurity
scattering. These effects have been further discussed in the
recent literature
\cite{Culcer2003,MURAKAMI2003A,Hu2003,Schliemann2003,Sinitsyn,Shen2003}.

Fundamental to the proposal of Murakami, Nagaosa and Zhang
\cite{MURAKAMI2003} is the spin-orbit coupling that exists in the
Luttinger effective-mass model in the degenerate valence bands:
\begin{equation}
H = \frac{1}{2m}  \left( (\gamma_1 + \frac{5}{2} \gamma_2) k^2 -
2\gamma_2 (\textbf{k} \cdot \textbf{S})^2 \right)
\end{equation}
\noindent In this "isotropic," or spherically symmetric model, the
helicity $\lambda = {\hat{k}} \cdot {\vec{S}}$ is a good quantum
number of the isotropic Luttinger hamiltonian above, and it labels
the two doubly degenerate Kramers bands corresponding to the heavy
holes $\lambda = \pm \frac{3}{2}$ and light holes $\lambda =\pm
\frac{1}{2}$ . The spin current effect can be intuitively
understood as a consequence of the conservation of total angular
momentum: $\textbf{J} = \hbar \textbf{x} \times \textbf{k} +
\textbf{S}$. The spin current flows in such a way that the change
of the orbital angular momentum $\textbf{L} = \hbar \textbf{x}
\times \textbf{k} $ exactly cancels the change of the spin angular
momentum $\textbf{S}$. When an electric field is applied on the
arbitrary $z$ axis, the $z$ component of $\textbf{J}$ is
conserved. The topological nature of the spin current is manifest
in the gauge-field formulation of \cite{MURAKAMI2003A}, where the
spin conductance is defined in terms of a linear combination of
the components of a gauge field, $G_{ij} = \lambda (\lambda^2
-13/4) \epsilon_{ijl} k_l /k^3$, clearly reflecting a monopole
structure in $\textbf{k}$ space. The singularity at $\textbf{k}
\longrightarrow 0$ exemplifies the confluence of the Kramers
doublets at the $\Gamma$ point where the band becomes $4$ fold
degenerate, but the flux of the gauge field through a $2$
dimensional surface in $\textbf{k}$ space is constant and set by
the helicity eigenvalue.

The picture presented above is valid as long as the Hamiltonian is
isotropic, that is to say it has spherical symmetry. In the real
materials in which the dissipationless spin-current is predicted
\cite{MURAKAMI2003}, all of which are characterized by a large
anisotropy (see Table~\ref{paramstable}), the angular momentum
$\textbf{J}$ and the helicity $\lambda = {\hat{k}} \cdot
{\vec{S}}$ are no longer good quantum numbers. It is therefore
vital to ask whether the topological spin current is preserved in
materials which are not rotationally invariant. In this paper, we
investigate the effect of the spherical symmetry breaking on the
dissipationless spin current, and calculate the values of the
intrinsic spin Hall conductivity for anisotropic band structure
parameters.

\begin{table}
 \begin{tabular}{|c|c|c|c|c|}
  \hline
  % after \\: \hline or \cline{col1-col2} \cline{col3-col4} ...
 & $\gamma_1$ & $\gamma_2$& $\gamma_3$ & $\delta$ \\
  \hline
  Si & 4.22 & 0.39 & 1.44 & 0.248 \\
  Ge & 13.35 & 4.25 & 5.69 & 0.108 \\
  GaAs & 4.15 & 1.01 & 1.75 & 0.114 \\
  InSb & 35.08 & 15.64 & 16.91 & 0.036 \\
  InAs & 19.67&8.37 &9.29 &0.047\\
  GaP & 4.20&0.98 &1.66 &0.162\\
  \hline
\end{tabular}
  \caption{Valence-band parameters for some common materials~\cite{LAWAETZ1971}. Following~\cite{BALDERES1973} we define
  $\delta = (\gamma_3-\gamma_2)/\gamma_1$ as a measure of the anisotropy.}\label{paramstable}
\end{table}

The most general Hamiltonian which respects time-reversal and
cubic symmetries was derived by Luttinger~\cite{luttinger1956}:
\begin{widetext}
\begin{multline}\label{hamiltonian}
H_0 = \frac{1}{2m} (\gamma_1 + \frac{5}{2} \gamma_2) k^2 -
\frac{\gamma_2}{m} (k_x^2S_x^2+ k_y^2S_y^2 + k_z^2S_z^2)-
2\frac{\gamma_3}{m}[\{k_x,k_y\}\{S_x,S_y\} +\{k_y,k_z\}
\{S_y,S_z\} +\{k_z,k_x\}\{S_z,S_x\}]
\end{multline}
\end{widetext}
where we define $\{A,B\} =\frac{1}{2}(AB+BA)$, and $k^2=k_x^2+
k_y^2+k_z^2$.  The parameters, $\gamma_1,\gamma_2,$ and
$\gamma_3,$ are material-dependent. In the special case of
$\gamma_2=\gamma_3$ (which we call isotropic), the last two terms
simply combine to yield $-\frac{\gamma_2}{m}(\vec k \cdot
\vec{S})^2$.

In real materials, however, the values of $\gamma_2$ and
$\gamma_3$ are very different. Table 1 lists the values of these
parameters in some important materials. The anisotropy,
characterized by the parameter $\delta\equiv
(\gamma_3-\gamma_2)/\gamma_1$, is relevant and substantial for all
the materials, and especially relevant for Si. In order to
understand the dissipationless spin-current generated in these
real materials, including its dependence on the orientation of the
field and current with respect to the crystal axes, we must
consider the full anisotropic Hamiltonian, Eq.~\ref{hamiltonian}.

When $\gamma_2\neq\gamma_3$, the Hamiltonian is no longer
isotropic and the helicity is not a good quantum number. However,
the energy spectrum of the Hamiltonian retains the same structure
as in the isotropic case, albeit with a different dispersion
relation. After diagonalizing the Hamiltonian, we obtain two
doubly degenerate energy levels, which we call light and heavy
holes in analogy with the isotropic case:
\begin{equation}
\begin{aligned}
 E(k)=& \frac{1}{2m} \gamma_1 k^2 \pm
 \frac{\gamma_3}{m} d(k) \\
 d^2(k)=&(\frac{\gamma_2}{\gamma_3})^2(k_x^4+k_y^4+k_z^4)\\
  &+(3-
 (\frac{\gamma_2}{\gamma_3})^2)(k_y^2 k_x^2 + k_x^2k_z^2 +k_y^2k_z^2).
\end{aligned}
\end{equation}
\noindent Following Ref. \cite{MURAKAMI2003A}, we can expand the
spin-dependent terms in the anisotropic Luttinger Hamiltonian in
terms of a Clifford algebra of Dirac $\Gamma$ matrices
$\{\Gamma^a, \Gamma^b \} = 2 \delta_{ab} I_{4 \times 4}$:
\begin{eqnarray}
H_0=\epsilon (\textbf{k}) + \frac{\gamma_3}{m}d_a\Gamma^a
\label{Clifford}
\end{eqnarray}
\begin{equation}
\begin{aligned}
\epsilon (\textbf{k}) & = \frac{\gamma_1}{2m} k^2, \\ d_1 & = -
\sqrt{3} k_z k_y, \; d_2 = - \sqrt{3} k_x k_z, \;
d_3 = - \sqrt{3}  k_x k_y,\\
d_4 &= - \frac{\sqrt{3}}{2} \frac{\gamma_2}{\gamma_3} (k_x^2 -
k_y^2), \; d_5 =- \frac{1}{2} \frac{\gamma_2}{\gamma_3}(2k_z^2 -
k_x^2 - k_y^2)
\end{aligned}
\end{equation}
with $d_ad_a=d^2$. Whereas in the isotropic Luttinger model the
matrix used to diagonalize the Hamiltonian belongs to the $SO(3)$
group of rotations in $\textbf{k}$ space \cite{MURAKAMI2003}, in
the anisotropic materials the matrix that diagonalizes the
anisotropic Hamiltonian belongs to the $SO(5)$ rotations in $d_a$
space. The $SO(5)$ Clifford algebra representation of the
Hamiltonian (\ref{Clifford}) naturally unifies both the isotropic
and the anisotropic Luttinger model on the same footing. Since
this form of the Hamiltonian depends on $\textbf{k}$ only through
the five dimensional vector $d_a$, a large part of the results in
\cite{MURAKAMI2003A} is directly applicable to the anisotropic
case. In this sense, the $SO(5)$ Clifford algebra formalism shows
its full power in the anisotropic case studied here. The
projection operators onto the two-dimensional subspace of states
of the heavy-hole (HH) and light-hole (LH) bands read:
\begin{equation}
P^L = \frac{1}{2} (1+ \hat{d_a} \Gamma^a) \;\;, \;\; P^H =
\frac{1}{2} (1- \hat{d_a} \Gamma^a)
\end{equation}
\noindent For finite $\textbf{k}$, the Hamiltonian maintains the
$SO(4)$ symmetry observed in \cite{MURAKAMI2003A}. This symmetry
reflects the degeneracy of the two Kramers doublets at each value
of $\textbf{k}$, corresponding to the doubly-degenerate HH and the
LH bands. Each of the bands has an $SU(2)$ symmetry, which we may
denote by $SU(2)_{HH}$ and $SU(2)_{LH}$. Therefore, the total
symmetry is $SU(2)_{HH} \times SU(2)_{LH} = SO(4)$. At the
$\Gamma$ point, $\textbf{k}=0$, there is an enhanced $SO(5)$
symmetry.

The symmetry generators read:
\begin{equation}
\begin{aligned}
\rho^{ab} = \Gamma^{ab} +  d_b d_c \Gamma^{ca} - d_a d_c
\Gamma^{cb} = \\
 = P^L \Gamma^{ab} P^L + P^H \Gamma^{ab} P^H
\end{aligned}
\end{equation}
\noindent where $\Gamma^{ab} = -\frac{i}{2} [\Gamma^a, \Gamma^b]$
and $[\rho^{ab}, H_0] =0$ trivially since the Hamiltonian is
diagonal in the HH and LH bands. The spin operators $S^i$ are
related to the $\Gamma^{ab}$ matrices through the tensor
$\eta^{i}_{ab},$ whose entries were given in \cite{MURAKAMI2003A}:
$S^i = \eta^{i}_{ab} \Gamma^{ab}$. The concept of a conserved spin
current is still valid in anisotropic materials, since the
projected spin is a constant of motion in virtue of its being a
linear combination of the symmetry generators, $S^l_{(c)} =
  \eta^{l}_{ab} \rho^{ab} =P^L S^l P^L + P^H S^l P^H$.
We can therefore define the conserved spin-current as $J^l_i =
\frac{1}{2} \left\{ \frac{\partial H}{\partial k_i} , S^l_{(c)}
\right\}$. Note that the richer anisotropic Luttinger Hamiltonian
yields a very similar structure to the isotropic one when cast in
$SO(4)$ language.

Although the concept of helicity $\lambda = k_i S^i$ is not valid
in anisotropic materials, we can define a corresponding conserved
helicity, $\lambda_{new}$, as:
\begin{equation}
\begin{aligned}
\lambda_{new} =k_i S^i_{(c)} = \lambda + 2k_i \eta^i_{ab}  d_b d_c
\Gamma^{ca} = \\ = P^L \lambda P^L + P^H \lambda P^H
\end{aligned}
\end{equation}
\noindent Since it is a linear combination of the symmetry
generators of $H_0$ ($\lambda_{new} =k_i S^i_{(c)} = k_i
\eta^i_{ab} \rho^{ab}$), it is clear that $[H, \lambda_{new}] =0$.
In the isotropic limit, $\lambda_{new} =\lambda$, as can be seen
using the identities $[\lambda, P^L] = [\lambda, P^H] = 0$, valid
in the isotropic case.

The recent work of Ref.\cite{MURAKAMI2003A} shows that the Kubo
formula for the conserved spin current response can be expressed
purely in terms of a geometric quantity
\begin{equation}
G_{ij} = G^{ab}_{ij} \Gamma^{ab} = \frac{1}{4d^3} \epsilon_{abcde}
d_c \frac{\partial d_d}{\partial k_i} \frac{\partial d_e}{\partial
k_j} \Gamma^{ab}
\end{equation}
\noindent which describes the mapping from the 3D $k$ vector space
to the 5D $d$ vector space. This results also includes a quantum
correction to the semiclassical result of Ref.\cite{MURAKAMI2003}.
We shall apply this formula to the anisotropic case here. However,
there is one essential difference. Whereas in the isotropic case,
the field strength can be brought, through a proper choice of
gauge, to the diagonal form $G_{ij} = \lambda (\lambda^2 -13/4)
\epsilon_{ijl} k_l /k^3$, in the anisotropic case this is
impossible. Non-abelian field strength are, in general,
gauge-variant. However, there is a fundamental difference between
fields that can be diagonalized through a gauge transformation and
fields for which this is not possible. The former are ultimately
abelian in nature, whereas the latter are truly non-abelian. The
non-diagonal gauge field which describes evolution in anisotropic
materials reflects the richer structure of the anisotropic
Luttinger hamiltonian.

We can express the field strength in terms of the (unprojected)
spin degrees of freedom if we first note that the ten $SO(5)$
generators $\Gamma^{ab}$ decompose into the $3$ spin matrices
$S^i$ and the seven cubic, symmetric and traceless combinations of
the spin operators of the form $S^i S^j S^k$, namely:
\begin{equation}
\begin{array}{c}
  A^1= (S_x)^3 , \; \;
  A^2 = (S_y)^3, \;\;
  A^3 = (S_z)^3 \\
  A^4 = \{S_x, (S_y)^2 - (S_z)^2 \} \\
  A^5 = \{S_y, (S_z)^2 - (S_x)^2 \}  \\
  A^6 = \{S_z, (S_x)^2 - (S_y)^2 \} \\
  A^7 = S_xS_yS_z + S_z S_y S_x \\
\end{array}
\end{equation}
\noindent Then we can write:
\begin{equation}
G_{ij} = \frac{1}{4d^3} \epsilon_{ijl} k_l [V_\mu A^\mu+ U_l S^l]
, \;\;\; l =1,...,3, \mu =1,...,7
\end{equation}
\noindent Where
\begin{equation}
\begin{array}{c}
U_l = \frac{1}{2} \frac{\gamma_2}{\gamma_3} [(13+28 \frac{\gamma_2}{\gamma_3}) k_l^3 + (13-28 \frac{\gamma_2}{\gamma_3})k^2 k_l] \\
  V_l = - 2 \frac{\gamma_2}{\gamma_3}[(1+4 \frac{\gamma_2}{\gamma_3}) k_l^3 + (1-4  \frac{\gamma_2}{\gamma_3})k^2 k_l] \\
  V_4 = -3 \frac{\gamma_2}{\gamma_3}  k_x(k_y^2 -k_z^2) \\
   V_5 = -3  \frac{\gamma_2}{\gamma_3} k_y(k_x^2 -k_z^2) \\
   V_6 = -3  \frac{\gamma_2}{\gamma_3} k_z(k_x^2 -k_y^2) \\
  V_7 = -12 k_x k_y k_z \\
\end{array}
\end{equation}
\noindent $l=1,..,3$

When cast in the $SO(4)$ language, the expression for the spin
conductance in anisotropic materials has the same form as in the
spherical model:
\begin{equation}
\sigma^{l}_{ij} = \frac{8e^2}{V\hbar} \sum_{k} (n_L (\textbf{k}) -
n_H(\textbf{k})) \frac{1}{3} \eta^{l}_{ab} G^{ab}_{ij}
\end{equation}
\noindent where $n_L = n_F(\epsilon_L)$ and $n_H=n_F(\epsilon_H)$
are the Fermi functions of the LH and HH bands.  This expression
can be put into the following elegant form:
\begin{equation}
\frac{1}{3} \eta^{l}_{ab} G^{ab}_{ij} = \frac{1}{8 d^3}
\frac{\gamma_2}{\gamma_3} \epsilon_{ijm} k_m  k_l
[(1-\frac{\gamma_2}{\gamma_3}) k_l^2 +
(1+\frac{\gamma_2}{\gamma_3}) k^2 ] \label{nosum}
\end{equation}
\noindent where we see that the first term in brackets vanishes in
the isotropic case. The $l$ index specifies the direction of the
spin orientation, and it is not summed over on the right hand side
of Eq. (\ref{nosum}). It is now obvious that the only components
of $\sigma^{l}_{ij}$ surviving after summing the contributions
from the whole Fermi surface, are those for which $i \ne j \ne l$.
Indeed, upon integration over $\textbf{k}$, $\sigma^{l}_{ij}$
becomes proportional to $\epsilon_{ijk}$, just as it should for in
crystals with cubic symmetry \cite{Lax}.

Our result for the spin-current can thus be put in the form:
\begin{equation}
\sigma_s=\frac{e^2}{\hbar}n^{1/3}S(\gamma_1,\gamma_2,\gamma_3),
\end{equation}
\noindent where the material-specific coefficient, $S$, is
independent of the Fermi energy, and is of the order $\sim 0.05$
for most materials (see Table~\ref{sigmatable}). The $\sigma_s
\sim n^{1/3}$ scaling is the hallmark of the dissipationless spin
current, and has been proposed as a means to distinguish from
other extrinsic effects\cite{MURAKAMI2003,Culcer2003}. To compare
the spin-conductance in different materials, we separate the
dependence on the total carrier density, which for the anisotropic
Luttinger model depends on the band-parameters:
$n=(2m\varepsilon_F)^{3/2}\frac{2}{3}\int\left[\frac{1}{(\gamma_1-2\gamma_3
d/k^2)^{3/2}}+\frac{1}{(\gamma_1+2\gamma_3
d/k^2)^{3/2}}\right]\frac{d^2\hat{k}}{(2\pi)^3}$. Using this
relation, we can find $\varepsilon_F$ as a function of $n$, and
use it to define the anisotropic Fermi distribution functions,
$n_{L,H}(\vk)=\Theta(k_F^{L,H}-k)$.  We have calculated $\sigma_s$
for band parameters corresponding to a selection of real
materials, as well as for band parameters corresponding to
isotropic materials with the same values of $\gamma_1$ and
$\mu\equiv \frac{6\gamma_3+4\gamma_2}{5\gamma_1}$.  The results,
listed in Table~\ref{sigmatable}, show that the non-zero
anisotropy leads to a decrease in the spin-conductivity of as much
as $30\%$ (for Si), although the reduction in materials with
smaller anisotropy is typically only $\sim 5\%$.

\begin{figure}
  % Requires \usepackage{graphicx}
  \includegraphics[width=2.5in]{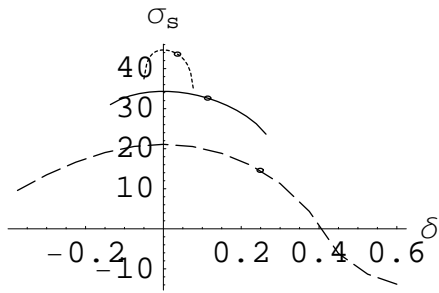}\\
  \caption{Spin-conductivity is plotted as a function of the anisotropy, parameterized by $\delta\equiv
  (\gamma_2-\gamma_3)/\gamma_1,$ with $\mu=
  (6\gamma_3+4\gamma_2)/5$ and $n=10^{19}$cm$^{-3}$ held fixed at
  values corresponding to Si (bottom curve), GaAs and InSb (top curve).
  The circles indicate the real values of the parameters in Si, GaAs and InSb.}\label{deltplot}
\end{figure}

\begin{figure}
  % Requires \usepackage{graphicx}
  \includegraphics[width=2.5in]{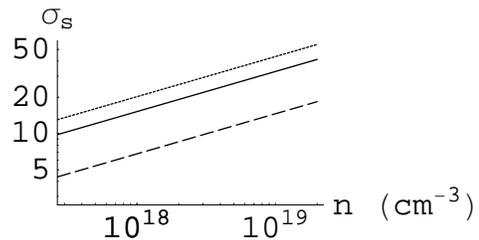}\\
  \caption{The dependence of spin-conductivity on carrier density, $n$, is plotted using the
  band parameters of Si (bottom curve), GaAs and InSb (top curve).}\label{plotn}
\end{figure}

To illustrate the systematic dependence of spin-conductance on
anisotropy, we plot $\sigma_s$ as a function of
$\delta=(\gamma_3-\gamma_2)/\gamma_1$ with $\gamma_1$ and $\mu=
(6\gamma_3+4\gamma_2)/5$ held fixed at the values corresponding to
Si, GaAs and InSb (Figure~\ref{deltplot}).  The spin-conductance
at fixed carrier concentration is maximum at $\delta=0$, whereas
all real materials have $\delta>0$. This observation should guide
the selection of materials with relatively low anisotropy for
spin-injection devices and other applications where a stron
spin-current is desired.  Finally, the variation of $\sigma_s$
with carrier concentration, $n$, is shown in Figure~\ref{plotn}.

We would like to thank L.~Balents, S.~Murakami, N.~Nagaosa and
J.~Sinova for the useful conversations. This work is supported by
the NSF under grant numbers DMR-9814289 and the US Department of
Energy, Office of Basic Energy Sciences under contract
DE-AC03-76SF00515. A. Bernevig acknowledges support from the
Stanford Graduate Fellowship and E. Mukamel acknowledges support
from the NSF Graduate Fellowship. JP is supported by the funds
from the David Saxon chair at UCLA.

\begin{table}
 \begin{tabular}{|c|c|c|c|c|c|}
  \hline
  % after \\: \hline or \cline{col1-col2} \cline{col3-col4} ...
 & $S(\gamma_1,\gamma_2,\gamma_3)$ & $\sigma_s$ ($\Omega^{-1}$ cm$^{-1}$)& $\sigma_s|_{\delta=0}(\Omega^{-1}$cm$^{-1})$  \\
  \hline
  Si & 0.028&14.60& 21.10 \\
  Ge & 0.063&32.79& 34.31\\
  GaAs&0.062&32.64& 34.33\\
  InSb&0.083&43.63& 44.67\\
  InAs&0.079&41.61& 42.55\\
  GaP &0.051&26.50& 29.17\\
  \hline
\end{tabular}
  \caption{We list the material-dependent coefficients of the
  spin-conductivity for values of $\gamma_1,\gamma_2,\gamma_3$
  corresponding to common semiconductors.  Also given are the
  actual spin-conductivities at $n=10^{19}$cm$^{-3}$ for both the
  real anisotropic materials, and their spherical approximations ($\delta=0$).}\label{sigmatable}
\end{table}

\bibliography{reference1,reference2}

\end{document}